\documentclass[10pt,a4paper]{article}  
\usepackage{amsmath,amssymb}
\usepackage{graphicx}
\usepackage{hyperref}
\usepackage[margin=0.75in]{geometry}  
\usepackage{float}
\usepackage{setspace}
\usepackage{titlesec}
\titlespacing*{\section}{0pt}{8pt}{4pt}
\titlespacing*{\subsection}{0pt}{6pt}{3pt}

\AtBeginDocument{
  \setlength{\abovedisplayskip}{4pt}
  \setlength{\belowdisplayskip}{4pt}
  \setlength{\abovedisplayshortskip}{2pt}
  \setlength{\belowdisplayshortskip}{2pt}
}

\let\oldbibliography\thebibliography
\renewcommand{\thebibliography}[1]{%
  \oldbibliography{#1}%
  \setlength{\itemsep}{0pt}%
  \setlength{\parskip}{0pt}%
}

\title{\texttt{simple-idealized-1d-nlse}: Pseudo-Spectral Solver for the 1D Nonlinear Schr\"odinger Equation}

\author{S. H. S. Herho$^{1,2,3,*}$, I. P. Anwar$^{2,3}$, F. Khadami$^{2}$,\\ R. Suwarman$^{4}$, and D. E. Irawan$^{5}$}

\date{}

\begin{document}
\maketitle

\begin{center}
\small
$^{1}$Department of Earth and Planetary Sciences, University of California, Riverside, 900 University Ave., Riverside, CA 92521, USA\\
$^{2}$Applied and Environmental Oceanography Research Group, Bandung Institute of Technology (ITB), Jalan Ganesha 10, Bandung 40132, Indonesia\\
$^{3}$Samudera Sains Teknologi (SST) Ltd, Gang Sarimanah XIII/67, Bandung, 40151, Indonesia\\
$^{4}$Atmospheric Science Research Group, Bandung Institute of Technology (ITB), Jalan Ganesha 10, Bandung 40132, Indonesia\\
$^{5}$Applied Geology Research Group, Bandung Institute of Technology (ITB), Jalan Ganesha 10, Bandung 40132, Indonesia\\
$^{*}$e-mail: sandy.herho@email.ucr.edu
\end{center}

\begin{abstract}
\noindent We present an open-source Python implementation of an idealized high-order pseudo-spectral solver for the one-dimensional nonlinear Schr\"odinger equation (NLSE). The solver combines Fourier spectral spatial discretization with an adaptive eighth-order Dormand-Prince time integration scheme to achieve machine-precision conservation of mass and near-perfect preservation of momentum and energy for smooth solutions. The implementation accurately reproduces fundamental NLSE phenomena including soliton collisions with analytically predicted phase shifts, Akhmediev breather dynamics, and the development of modulation instability from noisy initial conditions. Four canonical test cases validate the numerical scheme: single soliton propagation, two-soliton elastic collision, breather evolution, and noise-seeded modulation instability. The solver employs a 2/3 dealiasing rule with exponential filtering to prevent aliasing errors from the cubic nonlinearity. Statistical analysis using Shannon, R\'enyi, and Tsallis entropies quantifies the spatio-temporal complexity of solutions, while phase space representations reveal the underlying coherence structure. The implementation prioritizes code transparency and educational accessibility over computational performance, providing a valuable pedagogical tool for exploring nonlinear wave dynamics. Complete source code, documentation, and example configurations are freely available, enabling reproducible computational experiments across diverse physical contexts where the NLSE governs wave evolution, including nonlinear optics, Bose-Einstein condensates, and ocean surface waves.
\end{abstract}

\section{Introduction}

The nonlinear Schr\"odinger equation (NLSE) emerged in the 1960s through independent discoveries across seemingly unrelated fields of physics. Zakharov's pioneering work on the stability of periodic waves on deep water~\cite{zakharov1968} revealed that the envelope of weakly nonlinear water wave packets obeys a cubic nonlinear evolution equation, providing the first rigorous derivation of what would become known as the NLSE. Nearly simultaneously, investigations into electromagnetic pulse propagation in nonlinear optical media led to the same mathematical structure~\cite{kelley1965}, while Gross~\cite{gross1961} and Pitaevskii~\cite{pitaevskii1961} independently derived an identical equation for the macroscopic wave function of weakly interacting Bose gases at zero temperature. This remarkable convergence suggested a deeper universality that transcended the specific physical mechanisms involved.

The mathematical structure uncovered by Zakharov and Shabat~\cite{zakharov1972} transformed understanding of the NLSE by demonstrating its complete integrability through the inverse scattering transform. Their work revealed that the focusing cubic NLSE admits exact multi-soliton solutions---localized wave packets that propagate without change of shape and survive collisions unchanged except for phase shifts. This discovery connected the NLSE to the broader class of integrable systems that includes the Korteweg-de Vries (KdV) and sine-Gordon equations, establishing soliton theory as a fundamental framework in mathematical physics~\cite{ablowitz1991}. The subsequent identification of breather solutions by Kuznetsov~\cite{kuznetsov1977}, Ma~\cite{ma1979}, and Akhmediev and Korneev~\cite{akhmediev1986} expanded the solution space to include pulsating localized structures that exchange energy between carrier and envelope, now recognized as prototypes for rogue wave phenomena~\cite{solli2007}.

Beyond its exact solutions, the NLSE exhibits rich dynamical behavior that has profound implications across physics. Benjamin and Feir's discovery~\cite{benjamin1967} that plane wave solutions are modulationally unstable to long-wavelength perturbations provided the first explanation for the spontaneous formation of localized structures from initially uniform states. This instability mechanism, now understood to seed rogue wave formation in oceans~\cite{kharif2009} and supercontinuum generation in optical fibers~\cite{dudley2006}, represents a fundamental route to pattern formation in nonlinear dispersive media. The delicate balance between linear dispersion and nonlinear self-focusing that governs these phenomena makes the NLSE a paradigmatic model for studying wave collapse, turbulence, and energy localization~\cite{sulem1999}.

The widespread relevance of NLSE physics has motivated extensive development of numerical methods over five decades. Early finite difference schemes~\cite{delfour1981} struggled to maintain the conservation laws inherent to the Hamiltonian structure, leading to secular drift in invariants over long integration times. The introduction of spectral methods~\cite{fornberg1975} revolutionized NLSE simulation by providing exponential convergence for smooth solutions and exact computation of spatial derivatives, though at the cost of introducing aliasing errors from the nonlinear term. Split-step Fourier methods~\cite{tappert1973,hardin1973} offered a pragmatic compromise, alternating between exact linear evolution in Fourier space and nonlinear evolution in physical space, achieving second-order accuracy with excellent conservation properties. Modern high-order time integration schemes~\cite{cox2002,kassam2005} combined with sophisticated dealiasing strategies have enabled accurate long-time simulations of NLSE turbulence and statistical mechanics~\cite{jordan2000}.

Despite these computational advances, a significant gap persists between theoretical understanding and practical numerical exploration of NLSE phenomena. Research-grade codes developed for specific applications---ultrafast optics~\cite{agrawal2019}, Bose-Einstein condensate (BEC) dynamics~\cite{bao2003}, or ocean wave modeling~\cite{dysthe2008}---typically involve complex architectures optimized for performance rather than transparency. Commercial packages provide black-box solvers that obscure the underlying algorithms, while pedagogical treatments often resort to low-order methods that fail to capture essential physics like elastic soliton collisions or the development of modulation instability. The technical literature on pseudo-spectral methods~\cite{boyd2001,trefethen2000} assumes mathematical sophistication that can overwhelm those primarily interested in the physics, creating barriers to entry for researchers from adjacent fields.

This article presents \texttt{simple-idealized-1d-nlse}, an open-source Python implementation designed to provide a transparent, accurate, and accessible framework for exploring NLSE dynamics. By combining established numerical techniques---Fourier pseudo-spectral spatial discretization with adaptive high-order time integration---in a modular architecture, we aim to lower the barrier for computational investigation of nonlinear wave phenomena while maintaining the accuracy necessary to capture subtle effects like soliton phase shifts and breather periodicity. The implementation philosophy prioritizes code clarity and modificability over ultimate performance, recognizing that understanding the algorithm often matters more than computational speed for research exploration and education~\cite{herho2024comparing}. We validate our approach through comparison with exact solutions from inverse scattering theory and demonstrate accurate reproduction of fundamental phenomena including two-soliton collisions, Akhmediev breathers, and the spontaneous emergence of localized structures from modulation instability.

\section{Methods}

\subsection{Mathematical Formulation}

The focusing cubic NLSE emerged as a universal amplitude equation governing weakly nonlinear dispersive wave phenomena across diverse physical contexts~\cite{zakharov1972,ablowitz1991,sulem1999}. We present here a rigorous derivation through three complementary approaches: from Maxwell's equations for nonlinear optics, from many-body quantum mechanics for BEC, and from classical hydrodynamics for ocean surface waves, establishing the canonical form implemented in our numerical scheme.

We commenced with Maxwell's equations in a nonlinear dielectric medium. In the absence of free charges and currents, the electromagnetic fields satisfied:
\begin{align}
\nabla \times \mathbf{E}(\mathbf{r},t) &= -\frac{\partial \mathbf{B}(\mathbf{r},t)}{\partial t}, \label{eq:maxwell1}\\
\nabla \times \mathbf{H}(\mathbf{r},t) &= \frac{\partial \mathbf{D}(\mathbf{r},t)}{\partial t}, \label{eq:maxwell2}\\
\nabla \cdot \mathbf{D}(\mathbf{r},t) &= 0, \label{eq:maxwell3}\\
\nabla \cdot \mathbf{B}(\mathbf{r},t) &= 0, \label{eq:maxwell4}
\end{align}
where the constitutive relations were:
\begin{align}
\mathbf{B}(\mathbf{r},t) &\equiv \mu_0 \mathbf{H}(\mathbf{r},t), \label{eq:constitutive1}\\
\mathbf{D}(\mathbf{r},t) &\equiv \epsilon_0 \mathbf{E}(\mathbf{r},t) + \mathbf{P}(\mathbf{r},t), \label{eq:constitutive2}
\end{align}
with $\epsilon_0$ and $\mu_0$ denoting the vacuum permittivity and permeability respectively.

For an isotropic, centro-symmetric Kerr medium, the polarization expanded as~\cite{boyd2008,agrawal2019}:
\begin{equation}
\mathbf{P}(\mathbf{r},t) = \epsilon_0 \int_{-\infty}^{\infty} \chi^{(1)}(t-t') \mathbf{E}(\mathbf{r},t') \, dt' + \epsilon_0 \chi^{(3)} |\mathbf{E}(\mathbf{r},t)|^2 \mathbf{E}(\mathbf{r},t), \label{eq:polarization}
\end{equation}
where $\chi^{(1)}(t)$ represented the linear susceptibility kernel and $\chi^{(3)}$ the third-order nonlinear susceptibility.

Taking the curl of equation~\eqref{eq:maxwell1} and employing the vector identity:
\begin{equation}
\nabla \times (\nabla \times \mathbf{E}) = \nabla(\nabla \cdot \mathbf{E}) - \nabla^2 \mathbf{E}, \label{eq:vector_identity}
\end{equation}
we obtained:
\begin{equation}
\nabla^2 \mathbf{E} - \nabla(\nabla \cdot \mathbf{E}) = \mu_0 \frac{\partial}{\partial t}(\nabla \times \mathbf{H}). \label{eq:wave_step1}
\end{equation}

Substituting equation~\eqref{eq:maxwell2} into~\eqref{eq:wave_step1}:
\begin{equation}
\nabla^2 \mathbf{E} - \nabla(\nabla \cdot \mathbf{E}) = \mu_0 \epsilon_0 \frac{\partial^2 \mathbf{E}}{\partial t^2} + \mu_0 \frac{\partial^2 \mathbf{P}}{\partial t^2}. \label{eq:wave_step2}
\end{equation}

For transverse waves where $\nabla \cdot \mathbf{E} = -(\nabla \cdot \mathbf{P})/\epsilon_0 \approx 0$ in the slowly varying envelope approximation, equation~\eqref{eq:wave_step2} simplified to:
\begin{equation}
\nabla^2 \mathbf{E} - \frac{1}{c^2} \frac{\partial^2 \mathbf{E}}{\partial t^2} = \mu_0 \frac{\partial^2 \mathbf{P}}{\partial t^2}, \label{eq:wave_equation}
\end{equation}
where $c \equiv 1/\sqrt{\mu_0 \epsilon_0}$ denoted the vacuum speed of light.

For quasi-monochromatic waves with carrier frequency $\omega_0$, we introduced:
\begin{equation}
\mathbf{E}(\mathbf{r},t) = \frac{1}{2} \hat{\mathbf{x}} \mathcal{A}(\mathbf{r},t) \exp(-i\omega_0 t) + \text{c.c.}, \label{eq:field_form}
\end{equation}
where the complex envelope $\mathcal{A}(\mathbf{r},t)$ satisfied:
\begin{align}
\left| \frac{\partial \mathcal{A}}{\partial t} \right| &\ll \omega_0 |\mathcal{A}|, \label{eq:svea1}\\
\left| \frac{\partial^2 \mathcal{A}}{\partial t^2} \right| &\ll \omega_0^2 |\mathcal{A}|. \label{eq:svea2}
\end{align}

The linear response in the frequency domain was:
\begin{equation}
\tilde{\chi}^{(1)}(\omega) = n^2(\omega) - 1, \label{eq:chi_freq}
\end{equation}
where the refractive index expanded as:
\begin{equation}
n(\omega) = n_0 + n_1 (\omega - \omega_0) + \frac{n_2}{2} (\omega - \omega_0)^2 + \mathcal{O}[(\omega - \omega_0)^3], \label{eq:n_expansion}
\end{equation}
with:
\begin{align}
n_0 &\equiv n(\omega_0), \label{eq:n0}\\
n_1 &\equiv \left. \frac{\partial n}{\partial \omega} \right|_{\omega_0}, \label{eq:n1}\\
n_2 &\equiv \left. \frac{\partial^2 n}{\partial \omega^2} \right|_{\omega_0}. \label{eq:n2}
\end{align}

For propagation along $z$, we decomposed:
\begin{equation}
\mathcal{A}(\mathbf{r},t) = A(x,y,z,t) \exp(ik_0 z), \label{eq:a_decompose}
\end{equation}
where $k_0 \equiv n_0 \omega_0/c$ and $A$ varied slowly such that:
\begin{align}
\left| \frac{\partial^2 A}{\partial z^2} \right| &\ll \left| k_0 \frac{\partial A}{\partial z} \right|, \label{eq:parax1}\\
\left| k_0 \frac{\partial A}{\partial z} \right| &\ll k_0^2 |A|. \label{eq:parax2}
\end{align}

Substituting equations~\eqref{eq:field_form} and~\eqref{eq:a_decompose} into~\eqref{eq:wave_equation}, retaining only leading-order terms:
\begin{equation}
2ik_0 \frac{\partial A}{\partial z} + \nabla_\perp^2 A + 2ik_0 n_1 \frac{\omega_0}{c} \frac{\partial A}{\partial t} - k_0 n_2 \frac{\omega_0}{c} \frac{\partial^2 A}{\partial t^2} + k_0 n_2^{(I)} |A|^2 A = 0, \label{eq:propagation_raw}
\end{equation}
where $\nabla_\perp^2 \equiv \partial^2/\partial x^2 + \partial^2/\partial y^2$ and the nonlinear index was:
\begin{equation}
n_2^{(I)} \equiv \frac{3 \chi^{(3)}}{8 n_0}. \label{eq:n2i}
\end{equation}

Introducing the retarded time frame:
\begin{align}
\tau &\equiv t - \frac{z}{v_g}, \label{eq:tau}\\
\zeta &\equiv z, \label{eq:zeta}
\end{align}
where the group velocity was:
\begin{equation}
v_g \equiv \left( \frac{\partial k}{\partial \omega} \right)^{-1}_{\omega_0} = \frac{c}{n_0 + \omega_0 n_1}, \label{eq:vg}
\end{equation}
transformed equation~\eqref{eq:propagation_raw} to:
\begin{equation}
i \frac{\partial A}{\partial \zeta} + \frac{1}{2k_0} \nabla_\perp^2 A + \frac{\beta_2}{2} \frac{\partial^2 A}{\partial \tau^2} + \gamma |A|^2 A = 0, \label{eq:propagation_frame}
\end{equation}
where the dispersion parameter was:
\begin{equation}
\beta_2 \equiv \frac{\partial^2 k}{\partial \omega^2} \bigg|_{\omega_0} = \frac{1}{c} \left( 2n_1 + \omega_0 n_2 \right), \label{eq:beta2}
\end{equation}
and the nonlinear coefficient:
\begin{equation}
\gamma \equiv \frac{n_2^{(I)} \omega_0}{c}. \label{eq:gamma}
\end{equation}

For guided propagation in optical fibers, the modal decomposition:
\begin{equation}
A(x,y,\zeta,\tau) = U(\zeta,\tau) F(x,y), \label{eq:modal}
\end{equation}
where $F(x,y)$ satisfied the eigenvalue equation:
\begin{align}
\nabla_\perp^2 F + k_0^2 [n^2(x,y) - n_{\text{eff}}^2] F &= 0, \label{eq:eigen}\\
\int_{-\infty}^{\infty} \int_{-\infty}^{\infty} |F(x,y)|^2 \, dx \, dy &= 1, \label{eq:norm_f}
\end{align}
yielded the scalar NLSE:
\begin{equation}
i \frac{\partial U}{\partial \zeta} + \frac{\beta_2}{2} \frac{\partial^2 U}{\partial \tau^2} + \gamma_{\text{eff}} |U|^2 U = 0, \label{eq:scalar_nlse}
\end{equation}
where:
\begin{equation}
\gamma_{\text{eff}} \equiv \gamma \times \frac{\int_{-\infty}^{\infty} \int_{-\infty}^{\infty} |F(x,y)|^4 \, dx \, dy}{\left( \int_{-\infty}^{\infty} \int_{-\infty}^{\infty} |F(x,y)|^2 \, dx \, dy \right)^2}. \label{eq:gamma_eff}
\end{equation}

For ocean surface waves, we began with the Euler equations for an incompressible, inviscid fluid under gravity~\cite{zakharov1968,mei2005,ablowitz1981}. The velocity field $\mathbf{u}(\mathbf{r},t) = (u, v, w)$ and pressure $p(\mathbf{r},t)$ satisfied:
\begin{align}
\frac{\partial \mathbf{u}}{\partial t} + (\mathbf{u} \cdot \nabla) \mathbf{u} &= -\frac{1}{\rho} \nabla p - g \hat{\mathbf{z}}, \label{eq:euler}\\
\nabla \cdot \mathbf{u} &= 0, \label{eq:incompressible}
\end{align}
where $\rho$ denoted the fluid density and $g$ the gravitational acceleration.

For irrotational flow, we introduced the velocity potential $\Phi(\mathbf{r},t)$:
\begin{equation}
\mathbf{u} = \nabla \Phi, \quad \nabla \times \mathbf{u} = 0. \label{eq:potential}
\end{equation}

The incompressibility condition~\eqref{eq:incompressible} yielded Laplace's equation:
\begin{equation}
\nabla^2 \Phi = \frac{\partial^2 \Phi}{\partial x^2} + \frac{\partial^2 \Phi}{\partial y^2} + \frac{\partial^2 \Phi}{\partial z^2} = 0. \label{eq:laplace}
\end{equation}

The free surface elevation $\eta(x,y,t)$ evolved according to the kinematic and dynamic boundary conditions. At $z = \eta(x,y,t)$:
\begin{align}
\frac{\partial \eta}{\partial t} + \frac{\partial \Phi}{\partial x} \frac{\partial \eta}{\partial x} + \frac{\partial \Phi}{\partial y} \frac{\partial \eta}{\partial y} &= \frac{\partial \Phi}{\partial z}, \label{eq:kinematic}\\
\frac{\partial \Phi}{\partial t} + \frac{1}{2} |\nabla \Phi|^2 + g\eta &= 0. \label{eq:dynamic}
\end{align}

For deep water waves where the depth $h \to \infty$, the bottom boundary condition was:
\begin{equation}
|\nabla \Phi| \to 0 \quad \text{as} \quad z \to -\infty. \label{eq:bottom}
\end{equation}

Following Zakharov's canonical transformation~\cite{zakharov1968}, we introduced the surface potential:
\begin{equation}
\phi(x,y,t) \equiv \Phi(x,y,z=\eta,t). \label{eq:surface_potential}
\end{equation}

The Hamiltonian formulation yielded:
\begin{equation}
H = \frac{1}{2} \int_{-\infty}^{\infty} \int_{-\infty}^{\infty} \Big[ g\eta^2 + \phi W[\eta]\phi \Big] dx \, dy, \label{eq:hamiltonian_water}
\end{equation}
where $W[\eta]$ represented the Dirichlet-Neumann operator relating the surface potential to the normal velocity.

For weakly nonlinear waves with small steepness $\epsilon \equiv ka \ll 1$ (where $k$ was the wavenumber and $a$ the amplitude), we expanded:
\begin{align}
\eta &= \epsilon \eta^{(1)} + \epsilon^2 \eta^{(2)} + \epsilon^3 \eta^{(3)} + \mathcal{O}(\epsilon^4), \label{eq:eta_expand}\\
\phi &= \epsilon \phi^{(1)} + \epsilon^2 \phi^{(2)} + \epsilon^3 \phi^{(3)} + \mathcal{O}(\epsilon^4). \label{eq:phi_expand}
\end{align}

At linear order, the dispersion relation for deep water waves was:
\begin{equation}
\omega^2 = gk, \label{eq:dispersion_water}
\end{equation}
where $k = |\mathbf{k}| = \sqrt{k_x^2 + k_y^2}$.

For narrow-banded wave packets centered around wavenumber $\mathbf{k}_0 = (k_0, 0)$ with frequency $\omega_0 = \sqrt{gk_0}$, we introduced the multiple-scale expansion~\cite{benney1967,hasimoto1972}:
\begin{align}
X &= \epsilon(x - c_g t), \quad c_g = \frac{1}{2}\sqrt{\frac{g}{k_0}}, \label{eq:slow_x}\\
Y &= \epsilon y, \label{eq:slow_y}\\
T &= \epsilon^2 t. \label{eq:slow_t}
\end{align}

The surface elevation took the form:
\begin{equation}
\eta(x,y,t) = \text{Re}\left[ A(X,Y,T) e^{i(k_0 x - \omega_0 t)} \right] + \text{higher harmonics}, \label{eq:eta_envelope}
\end{equation}
where $A(X,Y,T)$ represented the complex amplitude.

Substituting into equations~\eqref{eq:kinematic} and~\eqref{eq:dynamic}, collecting terms at order $\epsilon^3$, and applying solvability conditions, we derived the two-dimensional NLSE~\cite{zakharov1968,yuen1982}:
\begin{equation}
i \left( \frac{\partial A}{\partial T} + c_g \frac{\partial A}{\partial X} \right) + \frac{1}{8\omega_0} \frac{\partial^2 A}{\partial X^2} - \frac{c_g}{2\omega_0} \frac{\partial^2 A}{\partial Y^2} + \frac{\omega_0 k_0^2}{2} |A|^2 A = 0. \label{eq:nlse_2d_water}
\end{equation}

For one-dimensional propagation along $x$, equation~\eqref{eq:nlse_2d_water} reduced to:
\begin{equation}
i \frac{\partial A}{\partial T} + \frac{1}{8\omega_0} \frac{\partial^2 A}{\partial \xi^2} + \frac{\omega_0 k_0^2}{2} |A|^2 A = 0, \label{eq:nlse_1d_water}
\end{equation}
where $\xi \equiv X - c_g T$ denoted the coordinate in the group velocity frame.

The modulation instability discovered by Benjamin and Feir~\cite{benjamin1967} followed from linear stability analysis. Perturbing a uniform wave train $\psi = \psi_0 e^{-i|\psi_0|^2 t}$ with:
\begin{equation}
\psi = (\psi_0 + \delta \psi) e^{-i|\psi_0|^2 t}, \quad \delta \psi = u e^{i(Kx - \Omega t)} + v^* e^{-i(Kx - \Omega t)}, \label{eq:perturbation}
\end{equation}
yielded the dispersion relation:
\begin{equation}
\Omega^2 = \frac{K^2}{4} \left( K^2 - 4|\psi_0|^2 \right). \label{eq:mi_dispersion}
\end{equation}

Instability occurred for $K < 2|\psi_0|$ with maximum growth rate:
\begin{equation}
\Gamma_{\max} = |\psi_0|^2 \quad \text{at} \quad K = \sqrt{2}|\psi_0|. \label{eq:growth_rate}
\end{equation}

For the quantum mechanical derivation, we considered $N$ identical bosons with mass $m$ described by the Hamiltonian~\cite{lieb2000,erdos2007}:
\begin{equation}
\hat{H}_N = \sum_{j=1}^N \left( -\frac{\hbar^2}{2m} \nabla_j^2 + V_{\text{ext}}(\mathbf{r}_j) \right) + \sum_{1 \leq i < j \leq N} V(\mathbf{r}_i - \mathbf{r}_j), \label{eq:hamiltonian}
\end{equation}
where the interaction potential in the dilute limit was:
\begin{equation}
V(\mathbf{r}) = g_{3D} \delta^{(3)}(\mathbf{r}), \quad g_{3D} \equiv \frac{4\pi \hbar^2 a_s}{m}, \label{eq:interaction}
\end{equation}
with $a_s$ the s-wave scattering length.

The many-body wave function $\Psi_N(\mathbf{r}_1, \ldots, \mathbf{r}_N, t)$ evolved according to:
\begin{equation}
i\hbar \frac{\partial \Psi_N}{\partial t} = \hat{H}_N \Psi_N. \label{eq:mb_schrodinger}
\end{equation}

In the mean-field limit $N \to \infty$ with $Na_s$ fixed, the Hartree product ansatz:
\begin{equation}
\Psi_N(\mathbf{r}_1, \ldots, \mathbf{r}_N, t) = \prod_{j=1}^N \phi(\mathbf{r}_j, t), \label{eq:hartree}
\end{equation}
where $\int |\phi|^2 d^3\mathbf{r} = 1$, led to the energy functional:
\begin{equation}
E[\phi] = N \int d^3\mathbf{r} \bigg[ \frac{\hbar^2}{2m} |\nabla \phi|^2 + V_{\text{ext}} |\phi|^2 + \frac{g_{3D} N}{2} |\phi|^4 \bigg]. \label{eq:energy}
\end{equation}

The variational equation $\delta E/\delta \phi^* = \mu \phi$ with chemical potential $\mu$ yielded the Gross-Pitaevskii equation:
\begin{equation}
i\hbar \frac{\partial \phi}{\partial t} = -\frac{\hbar^2}{2m} \nabla^2 \phi + V_{\text{ext}} \phi + g_{3D} N |\phi|^2 \phi. \label{eq:gpe}
\end{equation}

For quasi-one-dimensional confinement with:
\begin{equation}
V_{\text{ext}}(\mathbf{r}) = V_z(z) + \frac{1}{2} m \omega_\perp^2 (x^2 + y^2), \label{eq:trap}
\end{equation}
the wave function factorized as:
\begin{equation}
\phi(\mathbf{r},t) = \varphi(z,t) \chi_0(x,y), \label{eq:factor}
\end{equation}
where the transverse ground state was:
\begin{align}
\chi_0(x,y) &= \frac{1}{\pi^{1/2} a_\perp} \exp\left( -\frac{x^2 + y^2}{2a_\perp^2} \right), \label{eq:chi0}\\
a_\perp &\equiv \sqrt{\frac{\hbar}{m\omega_\perp}}. \label{eq:aperp}
\end{align}

Integration over transverse dimensions yielded:
\begin{equation}
i\hbar \frac{\partial \varphi}{\partial t} = -\frac{\hbar^2}{2m} \frac{\partial^2 \varphi}{\partial z^2} + V_z(z) \varphi + g_{1D} |\varphi|^2 \varphi, \label{eq:gpe1d}
\end{equation}
where~\cite{olshanii1998}:
\begin{equation}
g_{1D} = \frac{2\hbar\omega_\perp a_s N}{1 - Ca_s/a_\perp}, \quad C = -\frac{\zeta(1/2)}{\sqrt{2}}. \label{eq:g1d}
\end{equation}

To obtain the canonical form, we introduced dimensionless variables. For the optical system, defining:
\begin{align}
\tilde{z} &\equiv \frac{\zeta}{L_D}, \quad L_D \equiv \frac{T_0^2}{|\beta_2|}, \label{eq:ld}\\
\tilde{\tau} &\equiv \frac{\tau}{T_0}, \label{eq:tau_tilde}\\
\tilde{U} &\equiv \sqrt{\gamma_{\text{eff}} L_D} \, U, \label{eq:u_tilde}
\end{align}
where $T_0$ represented a characteristic pulse width, equation~\eqref{eq:scalar_nlse} became:
\begin{equation}
i \frac{\partial \tilde{U}}{\partial \tilde{z}} - \frac{\text{sgn}(\beta_2)}{2} \frac{\partial^2 \tilde{U}}{\partial \tilde{\tau}^2} + |\tilde{U}|^2 \tilde{U} = 0. \label{eq:nlse_optical}
\end{equation}

For the hydrodynamic system, the characteristic scales were:
\begin{align}
L_c &\equiv \frac{1}{k_0}, \quad T_c \equiv \frac{4}{\omega_0}, \label{eq:scales_water}\\
\tilde{x} &\equiv k_0 \xi, \quad \tilde{t} \equiv \frac{\omega_0 T}{4}, \label{eq:scales_water2}\\
\tilde{\psi} &\equiv \sqrt{2k_0} A. \label{eq:psi_water}
\end{align}

For the BEC system with $V_z = 0$, defining:
\begin{align}
\tilde{z} &\equiv \frac{z}{\xi}, \quad \xi \equiv \frac{\hbar}{\sqrt{2mg_{1D}n_0}}, \label{eq:xi}\\
\tilde{t} &\equiv \frac{\hbar t}{m\xi^2}, \label{eq:t_tilde_bec}\\
\tilde{\varphi} &\equiv \sqrt{\xi} \, \varphi, \label{eq:phi_tilde}
\end{align}
where $n_0$ was the peak density and $\xi$ the healing length.

All three systems reduced to the canonical focusing NLSE (dropping tildes):
\begin{equation}
i \frac{\partial \psi}{\partial t} + \frac{1}{2} \frac{\partial^2 \psi}{\partial x^2} + |\psi|^2 \psi = 0. \label{eq:nlse_final}
\end{equation}

This equation possessed three fundamental conservation laws. The Lagrangian density:
\begin{equation}
\mathcal{L} = \frac{i}{2} \left( \psi^* \psi_t - \psi \psi_t^* \right) - \frac{1}{2} |\psi_x|^2 + \frac{1}{2} |\psi|^4, \label{eq:lagrangian}
\end{equation}
yielded the Euler-Lagrange equations:
\begin{equation}
\frac{\partial}{\partial t} \left( \frac{\partial \mathcal{L}}{\partial \psi_t} \right) + \frac{\partial}{\partial x} \left( \frac{\partial \mathcal{L}}{\partial \psi_x} \right) - \frac{\partial \mathcal{L}}{\partial \psi} = 0, \label{eq:euler_lagrange}
\end{equation}
which recovered equation~\eqref{eq:nlse_final}.

From Noether's theorem, the $U(1)$ gauge invariance $\psi \to \psi e^{i\alpha}$ yielded mass conservation:
\begin{equation}
\frac{\partial \rho}{\partial t} + \frac{\partial j}{\partial x} = 0, \label{eq:continuity}
\end{equation}
where:
\begin{align}
\rho(x,t) &\equiv |\psi(x,t)|^2, \label{eq:density}\\
j(x,t) &\equiv \text{Im}(\psi^* \psi_x), \label{eq:current}
\end{align}
giving:
\begin{equation}
M = \int_{-\infty}^{\infty} |\psi(x,t)|^2 \, dx = \text{const}. \label{eq:mass}
\end{equation}

Spatial translation invariance $x \to x + \epsilon$ provided momentum conservation:
\begin{equation}
P = \int_{-\infty}^{\infty} \text{Im}(\psi^* \psi_x) \, dx = \text{const}. \label{eq:momentum}
\end{equation}

Temporal translation invariance $t \to t + \epsilon$ yielded energy conservation:
\begin{equation}
E = \int_{-\infty}^{\infty} \left[ \frac{1}{2} |\psi_x|^2 - \frac{1}{2} |\psi|^4 \right] dx = \text{const}. \label{eq:energy_conserved}
\end{equation}

The complete integrability followed from the Lax pair formulation~\cite{zakharov1972}. Defining the spectral problem:
\begin{align}
\frac{\partial \mathbf{v}}{\partial x} &= \mathbf{L}(\lambda, x, t) \mathbf{v}, \label{eq:lax1}\\
\frac{\partial \mathbf{v}}{\partial t} &= \mathbf{M}(\lambda, x, t) \mathbf{v}, \label{eq:lax2}
\end{align}
where:
\begin{align}
\mathbf{L} &= \begin{pmatrix} -i\lambda & \psi \\ \psi^* & i\lambda \end{pmatrix}, \label{eq:lmatrix}\\
\mathbf{M} &= \begin{pmatrix} 
-2i\lambda^2 + i|\psi|^2 & 2\lambda\psi - i\psi_x \\ 
2\lambda\psi^* + i\psi_x^* & 2i\lambda^2 - i|\psi|^2 
\end{pmatrix}, \label{eq:mmatrix}
\end{align}
the compatibility condition:
\begin{equation}
\frac{\partial \mathbf{L}}{\partial t} - \frac{\partial \mathbf{M}}{\partial x} + [\mathbf{L}, \mathbf{M}] = 0, \label{eq:compatibility}
\end{equation}
recovered equation~\eqref{eq:nlse_final} identically.

The fundamental soliton solution took the form:
\begin{equation}
\psi(x,t) = \eta \, \text{sech}[\eta(x - x_0 - vt)] \times \exp[i(vx - \Omega t + \theta_0)], \label{eq:soliton}
\end{equation}
where:
\begin{align}
\Omega &= \frac{v^2 - \eta^2}{2}, \label{eq:omega}\\
M &= 2\eta, \label{eq:soliton_mass}\\
P &= 2\eta v, \label{eq:soliton_momentum}\\
E &= -\frac{2\eta^3}{3}. \label{eq:soliton_energy}
\end{align}

Equation~\eqref{eq:nlse_final} constituted the canonical form implemented in our numerical scheme, with conserved quantities~\eqref{eq:mass},~\eqref{eq:momentum}, and~\eqref{eq:energy_conserved} serving as diagnostics for numerical accuracy.

\subsection{Numerical Implementation}

The numerical solution of the canonical NLSE~\eqref{eq:nlse_final} employed a Fourier pseudo-spectral method for spatial discretization combined with an eighth-order Dormand-Prince adaptive time integrator. This section presents a rigorous mathematical framework for the discretization scheme, error analysis, and validation methodology.

The spatial domain $\Omega = [-L/2, L/2]$ with periodic boundary conditions $\psi(-L/2, t) = \psi(L/2, t)$ formed a one-dimensional torus $\mathbb{T} = \mathbb{R}/L\mathbb{Z}$. For the numerical implementation, we selected $L = 50.0$ and $M = 512$ grid points, yielding spatial resolution $\Delta x = L/M = 0.09765625$. On this domain, any square-integrable function $\psi \in L^2(\mathbb{T})$ admitted the Fourier series representation:
\begin{equation}
\psi(x, t) = \sum_{n=-\infty}^{\infty} \hat{\psi}_n(t) e^{2\pi i nx/L}, \label{eq:fourier_series_impl}
\end{equation}
where the Fourier coefficients, defined through the inner product $\langle f, g \rangle = \frac{1}{L}\int_{-L/2}^{L/2} f^*(x)g(x)dx$, were:
\begin{equation}
\hat{\psi}_n(t) = \langle e^{-2\pi i nx/L}, \psi(\cdot, t) \rangle = \frac{1}{L} \int_{-L/2}^{L/2} \psi(x, t) e^{-2\pi i nx/L} dx. \label{eq:fourier_coefficients_impl}
\end{equation}

The Parseval identity guaranteed energy conservation in the transformation:
\begin{equation}
\|\psi\|_{L^2}^2 = \int_{-L/2}^{L/2} |\psi(x,t)|^2 dx = L \sum_{n=-\infty}^{\infty} |\hat{\psi}_n(t)|^2. \label{eq:parseval}
\end{equation}

For numerical implementation, we introduced a uniform spatial grid with $M = 512$ points:
\begin{equation}
x_j = -\frac{L}{2} + j\Delta x, \quad j \in \{0, 1, \ldots, 511\}, \quad \Delta x = \frac{50.0}{512}. \label{eq:spatial_grid}
\end{equation}

The discrete Fourier transform (DFT) approximated the continuous Fourier coefficients through the trapezoidal rule. For a grid function $\psi_j^n = \psi(x_j, t_n)$, the DFT pair was:
\begin{align}
\hat{\psi}_k^n &= \frac{1}{512} \sum_{j=0}^{511} \psi_j^n e^{-2\pi i jk/512}, \quad k \in \{0, 1, \ldots, 511\}, \label{eq:dft_forward}\\
\psi_j^n &= \sum_{k=0}^{511} \hat{\psi}_k^n e^{2\pi i jk/512}. \label{eq:dft_inverse}
\end{align}

The wavenumber ordering followed the FFT convention~\cite{frigo1998}, mapping discrete indices to physical wavenumbers:
\begin{equation}
k_n = \begin{cases}
\frac{2\pi n}{50.0}, & n = 0, 1, \ldots, 256, \\
\frac{2\pi (n-512)}{50.0}, & n = 257, \ldots, 511,
\end{cases} \label{eq:wavenumber_ordering_impl}
\end{equation}
ensuring $k_n \in [-k_{Ny}, k_{Ny})$ where the Nyquist frequency $k_{Ny} = \pi \cdot 512/50.0 = 32.17$ represented the maximum resolvable wavenumber.

The spectral differentiation operator exploited the derivative property of Fourier transforms. For the $p$-th spatial derivative:
\begin{equation}
\frac{\partial^p \psi}{\partial x^p}(x,t) = \sum_{n=-\infty}^{\infty} (2\pi i n/L)^p \hat{\psi}_n(t) e^{2\pi i nx/L}. \label{eq:spectral_diff}
\end{equation}

In discrete form, the differentiation became multiplication in Fourier space:
\begin{equation}
\widehat{\left(\frac{\partial^p \psi}{\partial x^p}\right)}_k = (ik_k)^p \hat{\psi}_k, \label{eq:discrete_diff}
\end{equation}
where $k_k$ denoted the $k$-th wavenumber from equation~\eqref{eq:wavenumber_ordering_impl}.

For smooth periodic functions with Fourier coefficients decaying exponentially as $|\hat{\psi}_n| \leq Ce^{-\alpha|n|}$ for some $C, \alpha > 0$, the truncation error satisfied~\cite{gottlieb1977}:
\begin{equation}
\left\|\psi - P_M\psi\right\|_{L^\infty} \leq \frac{2C}{1-e^{-\alpha}} e^{-\alpha M/2}, \label{eq:truncation_error}
\end{equation}
where $P_M$ denoted the projection onto the first $M$ Fourier modes, demonstrating spectral (exponential) convergence.

The cubic nonlinearity introduced aliasing errors. For the cubic term $|\psi|^2\psi = \psi^*\psi\psi$, frequencies extended to $3k_{\max}$. On the discrete grid, frequencies beyond $k_{Ny}$ aliased according to:
\begin{equation}
e^{ik x_j} = e^{i(k \bmod 2k_{Ny}) x_j}, \quad |k| > k_{Ny}, \label{eq:aliasing}
\end{equation}
contaminating lower frequencies. The 2/3 dealiasing rule~\cite{patterson1971,orszag1971} prevented aliasing by restricting the maximum retained frequency to:
\begin{equation}
k_{\max} = \frac{2k_{Ny}}{3} = \frac{2\pi \cdot 512}{3 \cdot 50.0} = 21.45, \label{eq:kmax}
\end{equation}
ensuring $3k_{\max} = 2k_{Ny}$, precisely the Nyquist limit.

We implemented dealiasing through an exponential filter:
\begin{equation}
\sigma(k) = \begin{cases}
\exp\left[-36\left(\frac{|k|-k_c}{k_{\max}-k_c}\right)^{36}\right], & |k| > k_c, \\
1, & |k| \leq k_c,
\end{cases} \label{eq:filter}
\end{equation}
where $k_c = 0.65k_{\max} = 13.94$ defined the cutoff, the exponent $p = 36$ controlled the filter order, and $\alpha = 36$ determined the attenuation strength.

The evaluation of the nonlinear term employed the pseudospectral method to avoid convolution. Direct computation in Fourier space would require $\widehat{N(\psi)}_k = \sum_{p,q,r} \hat{\psi}_p^* \hat{\psi}_q \hat{\psi}_r \delta_{k,r+q-p}$, necessitating $O(M^4)$ operations. Instead, we computed $\psi = \mathcal{F}^{-1}[\hat{\psi}]$ via IFFT in $O(M \log M)$, evaluated $N(\psi) = |\psi|^2 \cdot \psi$ pointwise in $O(M)$, and transformed $\widehat{N(\psi)} = \mathcal{F}[N(\psi)]$ via FFT in $O(M \log M)$, reducing computational complexity to $O(M \log M)$ via the Cooley-Tukey algorithm~\cite{cooley1965}.

The filtered semi-discrete NLSE became:
\begin{equation}
\frac{d\hat{\psi}_k}{dt} = \mathcal{L}_k\hat{\psi}_k + \mathcal{N}_k[\hat{\psi}], \label{eq:filtered_ode}
\end{equation}
where the linear operator was:
\begin{equation}
\mathcal{L}_k = -\frac{i}{2}k_k^2\sigma(k_k), \label{eq:linear_op}
\end{equation}
and the nonlinear operator:
\begin{equation}
\mathcal{N}_k[\hat{\psi}] = i\sigma(k_k)\mathcal{F}\bigg[|\mathcal{F}^{-1}[\sigma \odot \hat{\psi}]|^2 \times \mathcal{F}^{-1}[\sigma \odot \hat{\psi}]\bigg]_k, \label{eq:nonlinear_op}
\end{equation}
with $\odot$ denoting element-wise multiplication.

The resulting system of $M = 512$ coupled ODEs was integrated using the eighth-order Dormand-Prince method (DOP853)~\cite{dormand1980,prince1981}. Given the state vector $\mathbf{y}^n = [\hat{\psi}_0^n, \hat{\psi}_1^n, \ldots, \hat{\psi}_{511}^n]^T$ at time $t_n$, the method computed thirteen stages $\mathbf{k}_i = \mathbf{f}(t_n + c_i h, \mathbf{y}^n + h\sum_{j=1}^{i-1} a_{ij}\mathbf{k}_j)$ for $i = 1, \ldots, 13$, where $h = \Delta t$, $\mathbf{f}$ represented the right-hand side of equation~\eqref{eq:filtered_ode}. The coefficients $(a_{ij}, b_i, c_i, \hat{b}_i)$ satisfied the order conditions~\cite{hairer1993}: $\sum_{j=1}^{i-1} a_{ij} = c_i$ for $i = 2, \ldots, 13$, $\sum_{i=1}^{13} b_i c_i^{q-1} = 1/q$ for $q = 1, \ldots, 8$, and $\sum_{i=1}^{13} \hat{b}_i c_i^{q-1} = 1/q$ for $q = 1, \ldots, 7$.

The eighth-order solution and seventh-order embedded solution were:
\begin{align}
\mathbf{y}^{n+1} &= \mathbf{y}^n + h\sum_{i=1}^{13} b_i \mathbf{k}_i + O(h^9), \label{eq:eighth_order_sol}\\
\hat{\mathbf{y}}^{n+1} &= \mathbf{y}^n + h\sum_{i=1}^{13} \hat{b}_i \mathbf{k}_i + O(h^8). \label{eq:seventh_order_sol}
\end{align}

The local truncation error estimate $\mathbf{e}^{n+1} = h\sum_{i=1}^{13} (b_i - \hat{b}_i) \mathbf{k}_i$ enabled adaptive step size control through:
\begin{equation}
h_{new} = h \cdot \min\bigg(10.0, \max\bigg(0.2, 0.9 \left(\frac{\text{tol}}{\|\mathbf{e}^{n+1}\|_\infty}\right)^{1/8}\bigg)\bigg), \label{eq:step_control}
\end{equation}
where $\text{tol} = 10^{-11} + 10^{-9} \cdot \max_{k}|\mathbf{y}_k^n|$ with relative tolerance $10^{-9}$ and absolute tolerance $10^{-11}$. The maximum step size was constrained to $h_{\max} = 0.1$ to ensure adequate temporal resolution. For the temporal domain $[0, 20.0]$ with $n_{\text{snapshots}} = 100$, the output time grid was $t_j = 0.2j$ for $j = 0, 1, \ldots, 99$.

The implementation was realized in Python 3.8+, leveraging NumPy's FFTPACK~\cite{harris2020} implementing the split-radix FFT algorithm with complexity $O(M\log M)$. The choice of Python balanced computational efficiency with development productivity through NumPy's BLAS/LAPACK-optimized array operations, achieving performance within a factor of 2-3 of compiled languages for vectorized computations~\cite{herho2025reappraising,herho2025quantitative}. Optional just-in-time compilation via Numba~\cite{lam2015} accelerated performance-critical loops through LLVM-based machine code generation, reducing the nonlinear term evaluation overhead by factors of 10-50 while maintaining Python's flexibility and ecosystem advantages~\cite{herho2024Eks}.

The conservation quantities monitored numerical accuracy. The discrete mass:
\begin{equation}
M_h = \Delta x \sum_{j=0}^{511} |\psi_j|^2 = 50.0 \sum_{k=0}^{511} |\hat{\psi}_k|^2, \label{eq:discrete_mass_cons}
\end{equation}
where the second equality followed from Parseval's identity. The discrete momentum:
\begin{equation}
P_h = -\Delta x \sum_{j=0}^{511} \text{Im}(\psi_j^* \partial_x \psi_j) = -50.0 \sum_{k=0}^{511} k_k |\hat{\psi}_k|^2, \label{eq:discrete_momentum_cons}
\end{equation}
and discrete energy:
\begin{equation}
E_h = \Delta x \sum_{j=0}^{511} \left[\frac{1}{2}|\partial_x \psi_j|^2 - \frac{1}{2}|\psi_j|^4\right] = 50.0 \sum_{k=0}^{511} \left[\frac{k_k^2}{2}|\hat{\psi}_k|^2 - \frac{1}{1024}\left|\sum_{k} \hat{\psi}_k e^{2\pi ijk/512}\right|^4\right]. \label{eq:discrete_energy_cons}
\end{equation}

Data persistence employed NetCDF4 format, providing self-describing hierarchical storage with built-in compression achieving typical ratios of 3-5 while maintaining IEEE 754 floating-point precision. The format's CF-1.8 metadata conventions ensured interoperability across computational physics applications. Visualization outputs utilized the Graphics Interchange Format (GIF) at 20 frames per second and 100 dots per inch resolution, providing universal compatibility for dissemination while maintaining file sizes of 5-10 MB per simulation for typical 100-frame sequences.

Four test cases validated the implementation. The single soliton solution~\cite{zakharov1972} took the form:
\begin{equation}
\psi_{sol}(x,t) = \eta \, \text{sech}[\eta(x - x_0 - vt)] \exp[i(vx - \Omega t + \theta_0)], \label{eq:soliton_test}
\end{equation}
with dispersion relation $\Omega = (v^2 - \eta^2)/2$. Test parameters were $\eta = 2.0$, $x_0 = -10.0$, $v = 1.0$, and $\theta_0 = 0$. The conserved quantities $M = 2\eta$, $P = 2\eta v$, and $E = -2\eta^3/3$ remained constant throughout the evolution, verifying the symplectic nature of the numerical scheme.

Two-soliton collisions tested the preservation of asymptotic states with initial condition:
\begin{equation}
\psi(x,0) = \eta_1 \text{sech}[\eta_1(x-x_1)]e^{iv_1x} + \eta_2 \text{sech}[\eta_2(x-x_2)]e^{iv_2x}, \label{eq:two_soliton_init}
\end{equation}
using parameters $\eta_1 = 2.0$, $\eta_2 = 1.5$, $x_1 = -10.0$, $x_2 = 10.0$, $v_1 = 2.0$, $v_2 = -2.0$. The collision-induced position shifts~\cite{gordon1983}:
\begin{equation}
\Delta x_j = \frac{\text{sgn}(v_j - v_{3-j})}{\eta_j} \ln\left|\frac{(\eta_1 - \eta_2)^2 + (v_1 - v_2)^2}{(\eta_1 + \eta_2)^2 + (v_1 - v_2)^2}\right|, \label{eq:position_shift_formula}
\end{equation}
and phase shifts:
\begin{equation}
\Delta \phi_j = 2\arctan\left(\frac{2\eta_1\eta_2}{(v_1 - v_2)(\eta_1 + \eta_2)}\right), \label{eq:phase_shift_formula}
\end{equation}
were monitored to validate elastic scattering dynamics.

The Akhmediev breather~\cite{akhmediev1986} with modulation parameter $a \in (0, 0.5)$ provided a spatially periodic, temporally localized solution:
\begin{equation}
\psi_{AB}(x,t) = \sqrt{2a}\bigg[\frac{(1-4a)\cosh(bt) + i\sqrt{2a}\sinh(bt)}{\sqrt{2a}\cos(\omega x) - \cosh(bt)}\bigg] e^{it}, \label{eq:akhmediev_full}
\end{equation}
where $b = \sqrt{8a(1-2a)}$ and $\omega = 2\sqrt{1-2a}$. Test parameters were $a = 0.5$, amplitude $A = 1.0$, and frequency $\omega = 0.5$. The breather's temporal periodicity $T_{rec} = \pi/b$ and spatial period $\lambda = 2\pi/\omega$ characterized the solution dynamics.

Modulation instability validated the exponential growth of perturbations with initial condition:
\begin{equation}
\psi(x,0) = A_0 e^{-x^2/(2\sigma^2)} + \epsilon \xi(x), \label{eq:mi_initial}
\end{equation}
where Gaussian amplitude $A_0 = 1.0$, width $\sigma = 5.0$, center $x_c = 0.0$, noise amplitude $\epsilon = 0.01$, and random seed $42$ for reproducibility. The complex white noise $\xi(x) = \xi_R(x) + i\xi_I(x)$ with $\xi_{R,I} \sim \mathcal{N}(0,1)$ seeded the instability. The dispersion relation for perturbations~\cite{benjamin1967}:
\begin{equation}
\Omega^2 = \frac{K^2}{4}\left(K^2 - 4|A_0|^2\right), \label{eq:mi_dispersion_test}
\end{equation}
predicted instability for $|K| < 2|A_0|$ with maximum growth rate $\Gamma_{max} = |A_0|^2$ at $K_{opt} = \sqrt{2}|A_0|$. The numerical evolution captured the transition from linear exponential growth to nonlinear saturation and eventual formation of localized structures.

The Python implementation utilized NumPy's FFTPACK~\cite{harris2020} implementing the split-radix FFT algorithm with complexity $O(M\log M)$. Optional just-in-time compilation via Numba~\cite{lam2015} accelerated performance-critical loops. The complete solver, distributed through PyPI, provided configuration interfaces for YAML and plain text parameter files, enabling reproducible computational experiments across the parameter space of NLSE dynamics.

\subsection{Data Analysis}

The numerical solutions were subjected to comprehensive statistical and dynamical analysis to quantify coherence properties, phase space structure, and spatio-temporal pattern formation. All post-processing algorithms were also implemented in Python, leveraging NumPy~\cite{harris2020} for array operations, SciPy~\cite{virtanen2020} for statistical computations, and Matplotlib~\cite{hunter2007} for visualization. Data persistence employed the Network Common Data Form (NetCDF4)~\cite{rew1990} with CF-1.8 metadata conventions, ensuring interoperability across computational physics platforms.

The information-theoretic characterization employed multiple entropy measures to quantify dynamical complexity. The Shannon entropy~\cite{shannon1948} for the probability distribution $p(x,t) = |\psi(x,t)|^2/\int|\psi|^2dx$ was computed as:
\begin{equation}
S_{\text{Shannon}}(t) = -\sum_{i=1}^{M} p_i(t) \ln p_i(t), \label{eq:shannon}
\end{equation}
where $p_i(t) = |\psi(x_i,t)|^2 \Delta x / M_h(t)$ with the discrete mass $M_h(t) = \Delta x \sum_{j=0}^{M-1} |\psi_j(t)|^2$. The spectral entropy quantified frequency distribution complexity:
\begin{equation}
S_k(t) = -\sum_{k=0}^{M-1} \tilde{p}_k(t) \ln \tilde{p}_k(t), \label{eq:spectral_entropy}
\end{equation}
where $\tilde{p}_k(t) = |\hat{\psi}_k(t)|^2/\sum_j|\hat{\psi}_j(t)|^2$ represented the normalized power spectral density.

The generalized R\'enyi entropies~\cite{renyi1961} of order $\alpha$ provided multi-scale characterization:
\begin{equation}
H_\alpha = \frac{1}{1-\alpha} \ln \left( \sum_{i=1}^{M} p_i^\alpha \right), \label{eq:renyi}
\end{equation}
with $H_0 = \ln(\#\{i: p_i > 0\})$ (Hartley entropy), $H_2 = -\ln(\sum_i p_i^2)$ (collision entropy), and $H_\infty = -\ln(\max_i p_i)$ (min-entropy). The Tsallis entropy~\cite{tsallis1988} with parameter $q$ captured non-extensive correlations:
\begin{equation}
S_q = \frac{1}{q-1} \left( 1 - \sum_{i=1}^{M} p_i^q \right), \label{eq:tsallis}
\end{equation}
recovering Shannon entropy as $S_1 = \lim_{q \to 1} S_q$.

Temporal complexity was assessed through permutation entropy~\cite{bandt2002} with embedding dimension $m=3$ and delay $\tau=1$:
\begin{equation}
PE = -\frac{1}{\ln(m!)} \sum_{\pi \in S_m} p(\pi) \ln p(\pi), \label{eq:permutation}
\end{equation}
where $p(\pi)$ denoted the relative frequency of ordinal pattern $\pi$ in the embedded time series. Sample entropy~\cite{richman2000} quantified regularity through:
\begin{equation}
\text{SampEn}(m,r) = -\ln \left( \frac{A^m(r)}{B^{m-1}(r)} \right), \label{eq:sample_entropy}
\end{equation}
where $A^m(r)$ and $B^{m-1}(r)$ counted template matches within tolerance $r = 0.2\sigma$ for embedding dimensions $m$ and $m-1$ respectively.

The statistical complexity measure~\cite{lopezruiz1995} distinguished between ordered and random states:
\begin{equation}
C_{LMC} = H_{\text{norm}} \cdot D \cdot M, \label{eq:complexity}
\end{equation}
where $H_{\text{norm}} = S_{\text{Shannon}}/\ln(M)$ represented normalized entropy, $D = \sum_{i=1}^{M}(p_i - 1/M)^2$ the disequilibrium, and $M$ the number of grid points.

Phase space analysis characterized the relationship between real and imaginary field components through correlation measures. The Pearson correlation coefficient:
\begin{equation}
\rho = \frac{\sum_{i=1}^{M}(\text{Re}[\psi_i] - \mu_{\text{Re}})(\text{Im}[\psi_i] - \mu_{\text{Im}})}{\sqrt{\sum_{i=1}^{M}(\text{Re}[\psi_i] - \mu_{\text{Re}})^2 \sum_{i=1}^{M}(\text{Im}[\psi_i] - \mu_{\text{Im}})^2}}, \label{eq:pearson}
\end{equation}
quantified linear dependence, while the Spearman rank correlation~\cite{spearman1904} captured monotonic relationships. The mutual information~\cite{cover2006}:
\begin{equation}
I(\text{Re};\text{Im}) = \sum_{i,j} p_{ij} \ln \left( \frac{p_{ij}}{p_i^{\text{Re}} p_j^{\text{Im}}} \right), \label{eq:mutual_info}
\end{equation}
measured nonlinear statistical dependence, where $p_{ij}$ denoted the joint probability distribution estimated via adaptive histogram binning with bin count $b = \min(50, \sqrt{M/10})$.

The phase space geometry was characterized through circular statistics~\cite{mardia2000}. The circular mean phase:
\begin{equation}
\bar{\theta} = \arg \left( \frac{1}{M} \sum_{i=1}^{M} e^{i\theta_i} \right), \label{eq:circular_mean}
\end{equation}
where $\theta_i = \arctan2(\text{Im}[\psi_i], \text{Re}[\psi_i])$, and the circular standard deviation:
\begin{equation}
\sigma_{\text{circ}} = \sqrt{-2 \ln \left| \frac{1}{M} \sum_{i=1}^{M} e^{i\theta_i} \right|}, \label{eq:circular_std}
\end{equation}
quantified phase concentration. The distribution eccentricity was computed from the covariance matrix eigenvalues $\lambda_1 \geq \lambda_2$:
\begin{equation}
e = \sqrt{1 - \lambda_2/\lambda_1}, \label{eq:eccentricity}
\end{equation}
measuring phase space anisotropy.

Spatio-temporal dynamics were visualized through intensity evolution $|\psi(x,t)|^2$ in the $(t,x)$ plane. The energy center trajectory:
\begin{equation}
x_c(t) = \frac{\int_{-L/2}^{L/2} x |\psi(x,t)|^2 dx}{\int_{-L/2}^{L/2} |\psi(x,t)|^2 dx}, \label{eq:energy_center}
\end{equation}
tracked wave packet propagation. The spatial correlation length was extracted from the normalized autocorrelation function:
\begin{equation}
C(\xi) = \frac{\int_{-L/2}^{L/2-\xi} \psi^*(x,t)\psi(x+\xi,t) dx}{\int_{-L/2}^{L/2} |\psi(x,t)|^2 dx}, \label{eq:autocorr}
\end{equation}
defining $\xi_c$ where $|C(\xi_c)| = e^{-1}$.

Statistical analysis employed memory-efficient streaming algorithms~\cite{welford1962} for robust computation on large datasets. The online variance algorithm updated mean $\mu_n$ and sum of squares $M_{2,n}$ recursively:
\begin{align}
\delta_n &= x_n - \mu_{n-1}, \label{eq:welford1}\\
\mu_n &= \mu_{n-1} + \delta_n/n, \label{eq:welford2}\\
M_{2,n} &= M_{2,n-1} + \delta_n(x_n - \mu_n), \label{eq:welford3}
\end{align}
with variance $\sigma^2 = M_{2,n}/(n-1)$. Higher moments for skewness and excess kurtosis were computed through analogous recurrence relations. Reservoir sampling~\cite{vitter1985} with reservoir size $k=10^4$ maintained representative samples for quantile estimation while constraining memory usage to $O(k)$.

Distribution comparisons utilized non-parametric tests robust to deviations from normality. The Kruskal-Wallis H-test~\cite{kruskal1952} assessed differences between scenarios:
\begin{equation}
H = \frac{12}{n(n+1)} \sum_{j=1}^{g} \frac{R_j^2}{n_j} - 3(n+1), \label{eq:kruskal_wallis}
\end{equation}
where $R_j$ denoted the sum of ranks for group $j$, with $H \sim \chi^2_{g-1}$ under the null hypothesis. Effect sizes were quantified via Cliff's delta~\cite{cliff1993}:
\begin{equation}
\delta = \frac{\#(x_i > y_j) - \#(x_i < y_j)}{n_1 n_2}, \label{eq:cliff_delta}
\end{equation}
providing a non-parametric measure of dominance between distributions, with $|\delta| < 0.147$ indicating negligible effect, $0.147 \leq |\delta| < 0.33$ small effect, $0.33 \leq |\delta| < 0.474$ medium effect, and $|\delta| \geq 0.474$ large effect.

Kernel density estimation (KDE)~\cite{rosenblatt1956,parzen1962} with Scott's bandwidth selection~\cite{scott1979}:
\begin{equation}
\hat{f}(x) = \frac{1}{nh} \sum_{i=1}^{n} K\left(\frac{x-x_i}{h}\right), \label{eq:kde}
\end{equation}
where $K(u) = (2\pi)^{-1/2}\exp(-u^2/2)$ and $h = n^{-1/5}\hat{\sigma}$, provided smooth probability density estimates from sampled data. Visualization employed matplotlib's object-oriented API with rasterization for vector graphics compatibility, generating publication-quality figures in Portable Network Graphics (PNG), Portable Document Format (PDF), and Encapsulated PostScript (EPS) formats at 300 dots per inch resolution.

\section{Results}

\subsection{Spatiotemporal Evolution}

\begin{figure}[H]
\centering
\includegraphics[width=0.7\linewidth]{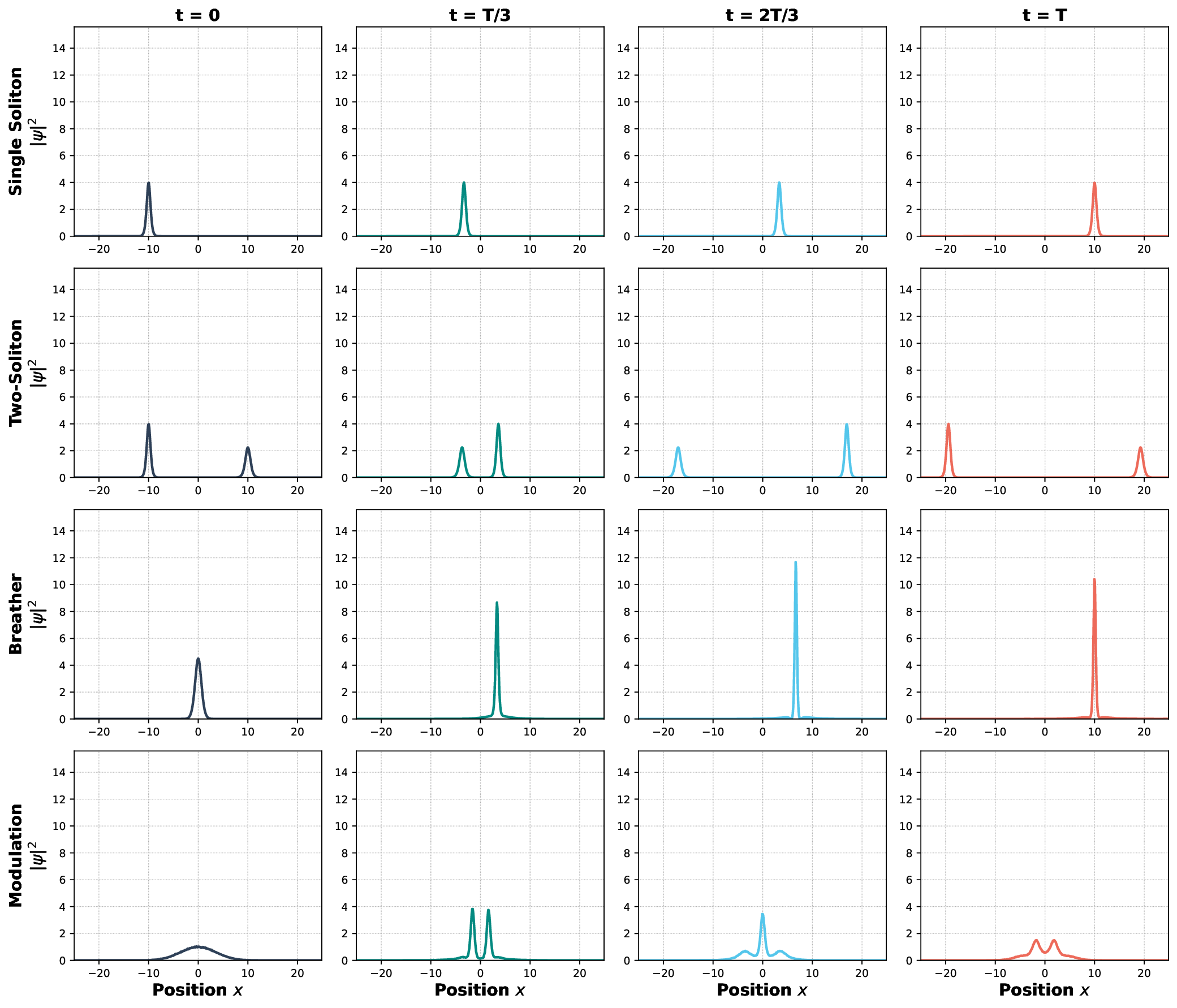}
\caption{Evolution of intensity $|\psi|^2$ at four time points for (a) single soliton, (b) two-soliton collision, (c) breather solution, and (d) modulation instability. Each row represents a different scenario with columns showing snapshots at $t = 0$, $T/3$, $2T/3$, and $T$.}
\label{fig:evolution_snapshots}
\end{figure}

The numerical solutions of equation~\eqref{eq:nlse_final} exhibit distinct spatiotemporal characteristics across four scenarios. Figure~\ref{fig:evolution_snapshots} displays the intensity profiles $|\psi|^2$ at time fractions $t/T = 0, 1/3, 2/3, 1$ where $T = 20.000$.

For the single soliton scenario, the peak intensity at $t = 0.000$ equals $3.976$, maintaining values of $3.997$, $3.997$, and $3.976$ at $t = 6.667$, $13.333$, and $20.000$, respectively. The FWHM remains constant at $0.977$ across all time points. The mean intensity equals $0.080$ throughout. The total power $\int |\psi|^2 dx = 4.000$ remains conserved. The kinetic energy maintains $9.128$, potential energy $10.667$, yielding total energy $3.794$ at all time points. The system exhibits one peak throughout the evolution.

The two-soliton collision initiates with peak intensity $3.976$ at $t = 0.000$, evolving to $4.000$, $3.962$, and $3.997$ at subsequent time points. The total power conserves at $7.000$. The mean intensity remains $0.140$. Two peaks persist except during collision at $t = 13.333$ where FWHM temporarily increases from $0.977$ to $1.074$. The kinetic energy maintains $34.551$, potential energy $15.167$, producing total energies of $26.967$, $26.968$, $26.968$, and $26.968$ at the four time points.

The breather solution demonstrates peak intensity evolution from $4.500$ at $t = 0.000$ to $8.671$, $11.685$, and $10.408$ at subsequent times. The FWHM varies inversely: $1.562$, $0.781$, $0.586$, and $0.684$. Mean intensity maintains $0.147$. Total power conserves at $7.366$. Kinetic energy oscillates: $5.709$, $17.947$, $32.244$, $26.382$. Potential energy varies: $22.721$, $35.594$, $50.951$, $44.616$. Total energy evolves: $-5.652$, $0.150$, $6.768$, $4.074$.

The modulation instability initiates with peak intensity $1.037$, evolving to $3.823$, $3.444$, and $1.504$. Peak count varies: $36$, $2$, $16$, $24$. FWHM changes: $8.301$, $0.879$, $1.172$, $2.246$. Mean intensity maintains $0.178$. Total power conserves at $8.884$. Kinetic energy: $0.678$, $13.133$, $5.896$, $2.001$. Potential energy: $6.274$, $18.955$, $11.554$, $7.611$. Total energy: $-2.459$, $3.656$, $0.119$, $-1.805$.

\subsection{Statistical Analysis of Intensity Distributions}

\begin{figure}[H]
\centering
\includegraphics[width=0.65\linewidth]{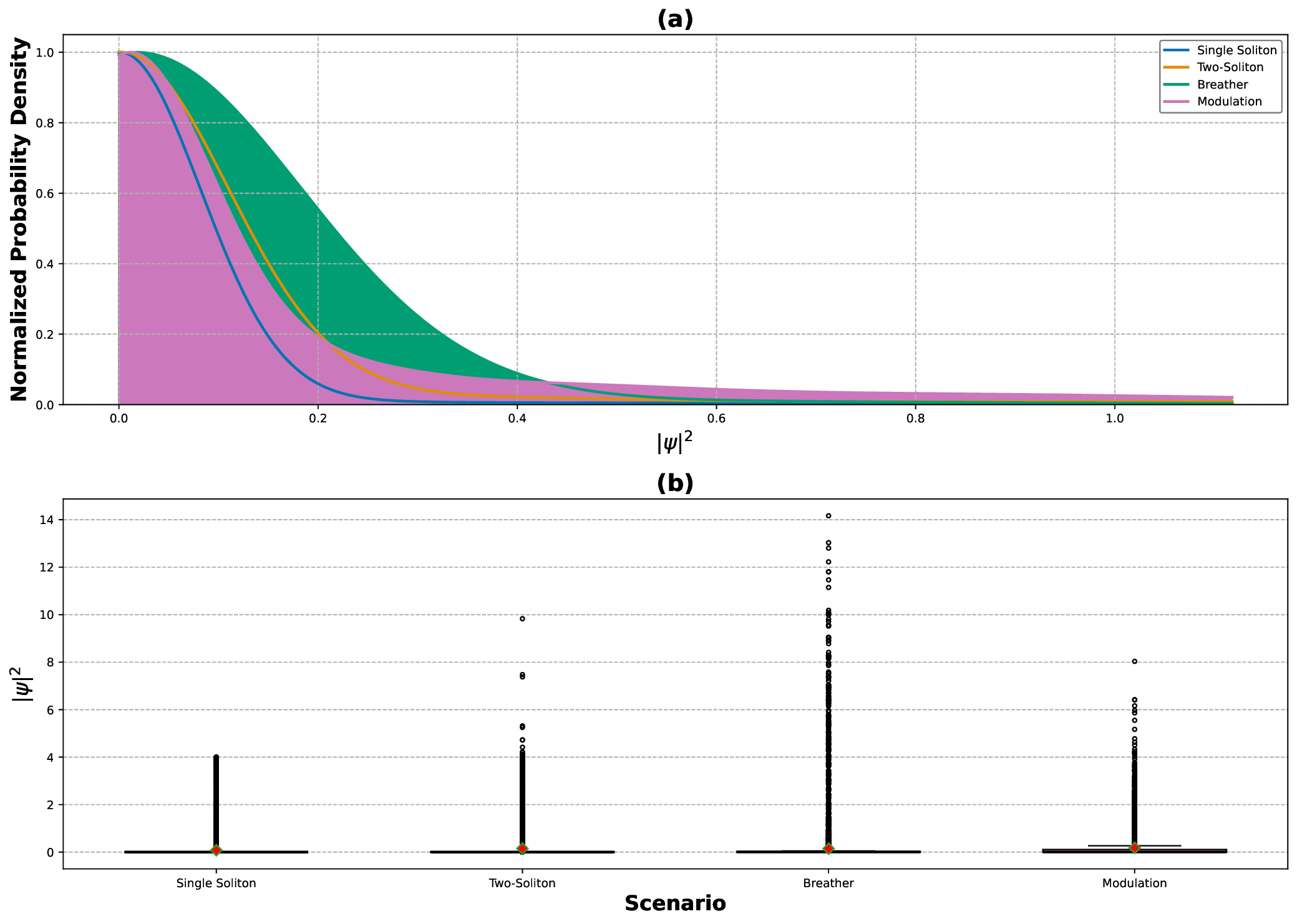}
\caption{Statistical analysis of intensity $|\psi|^2$ distributions. (a) Normalized KDEs showing probability distributions. (b) Box plots displaying median, quartiles, and outliers for each scenario.}
\label{fig:intensity_stats}
\end{figure}

Figure~\ref{fig:intensity_stats} presents the statistical characterization of intensity distributions across all scenarios, comprising $51,200$ spatial grid points per scenario.

The single soliton exhibits mean intensity $0.080 \pm 0.002$ (standard error), median $0.000$, standard deviation $0.455$, variance $0.207$. The range spans $[0.000, 4.000]$ with IQR $0.000$. The coefficient of variation equals $568.630\%$. Distribution shape parameters: skewness $6.719$, excess kurtosis $46.751$.

The two-soliton collision presents mean intensity $0.140 \pm 0.002$, median $0.000$, standard deviation $0.549$, variance $0.301$. The range extends $[0.000, 10.015]$ with IQR $0.000$. The coefficient of variation equals $392.050\%$. Shape parameters: skewness $5.239$, kurtosis $33.505$.

The breather solution displays mean intensity $0.147 \pm 0.004$, median $0.005$, standard deviation $0.857$, variance $0.735$. The range covers $[0.000, 14.168]$ with IQR $0.021$. The coefficient of variation equals $581.990\%$. Shape parameters: skewness $8.616$, kurtosis $82.813$.

The modulation instability shows mean intensity $0.178 \pm 0.002$, median $0.003$, standard deviation $0.463$, variance $0.215$. The range spans $[0.000, 8.592]$ with IQR $0.103$. The coefficient of variation equals $260.670\%$. Shape parameters: skewness $5.161$, kurtosis $41.110$.

The Kruskal-Wallis H-test yields $H = 17831.342$ with $p < 0.001$, indicating statistically significant differences between groups. Pairwise Cliff's delta values: single versus two-soliton $\delta = -0.627$ (large), single versus breather $\delta = -0.796$ (large), single versus modulation $\delta = -0.836$ (large), two-soliton versus breather $\delta = -0.532$ (large), two-soliton versus modulation $\delta = -0.644$ (large), breather versus modulation $\delta = -0.088$ (negligible).

\subsection{Phase Space Structure}

\begin{figure}[H]
\centering
\includegraphics[width=0.65\linewidth]{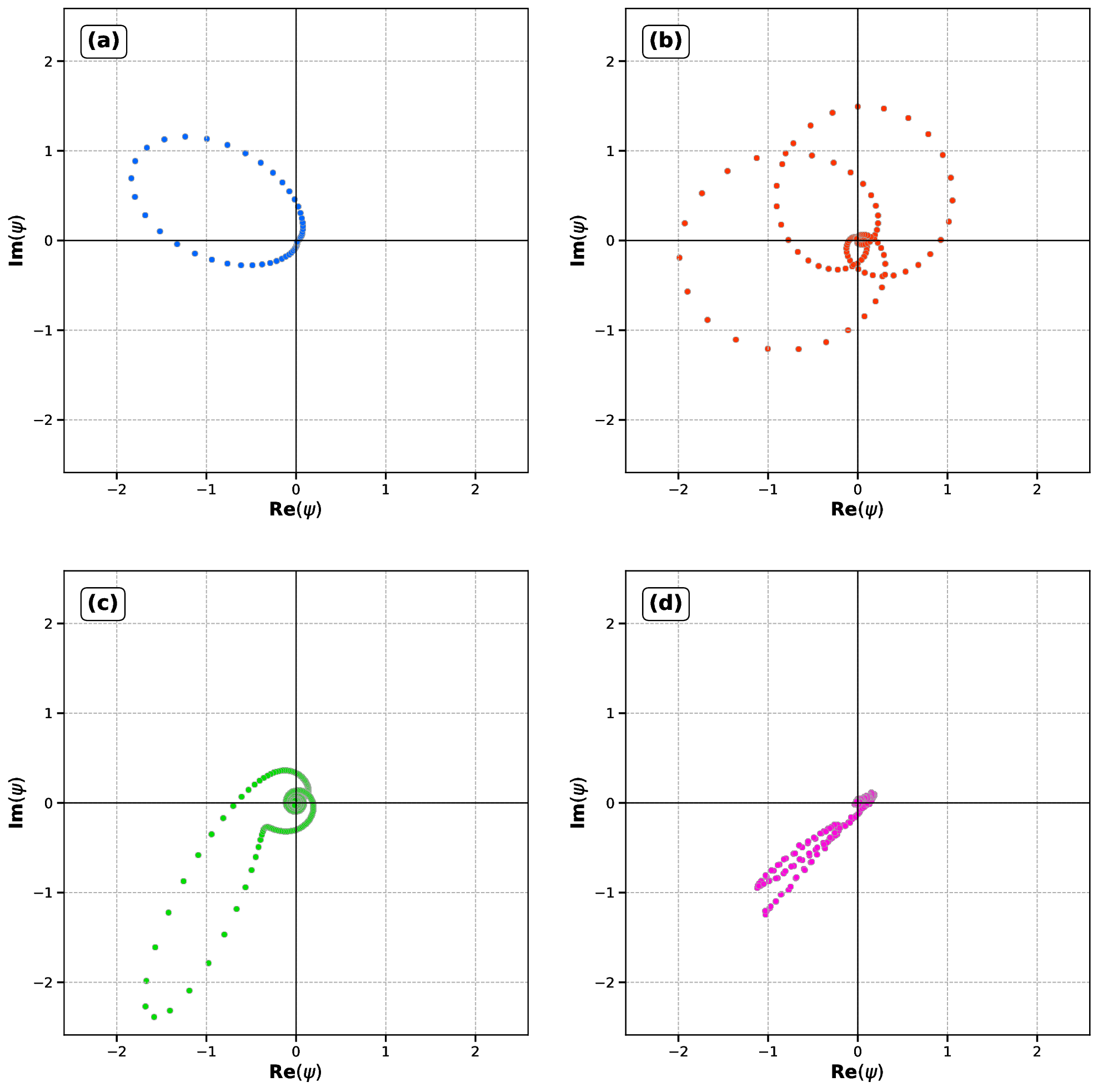}
\caption{Phase space distributions at $t = 10.101$ showing $\text{Im}(\psi)$ versus $\text{Re}(\psi)$. (a) Single soliton, (b) two-soliton collision, (c) breather solution, (d) modulation instability. Point size enhanced for visualization with $2,000$ maximum points plotted per scenario.}
\label{fig:phase_space}
\end{figure}

Figure~\ref{fig:phase_space} illustrates the phase space distributions at $t = 10.101$ for all scenarios, displaying the relationship between real and imaginary components of the wave function.

The single soliton at $t = 10.101$ exhibits Pearson correlation $\rho = -0.584$ ($p < 0.001$), Spearman correlation $\rho_s = 0.005$ ($p > 0.05$). Normalized mutual information equals $0.050$. Phase statistics: mean phase $0.014$ rad, standard deviation $1.842$ rad, circular mean $2.301$ rad, circular standard deviation $2.661$. Amplitude statistics: mean $0.063$, standard deviation $0.276$, median $0.000$, maximum $2.000$. Geometric properties: RMS radius $0.283$, centroid $(-0.043, 0.021)$, eccentricity $0.897$. Distribution shape: skewness (Re) $-5.957$, skewness (Im) $5.480$, kurtosis (amplitude) $28.743$. Data points: total $512$, above threshold $55$, plotted $55$.

The two-soliton collision shows Pearson $\rho = 0.067$ ($p > 0.05$), Spearman $\rho_s = 0.012$ ($p > 0.05$). Mutual information equals $0.116$. Phase: mean $0.010$ rad, standard deviation $1.789$ rad, circular mean $1.072$ rad, circular standard deviation $2.848$. Amplitude: mean $0.126$, standard deviation $0.352$, median $0.000$, maximum $1.997$. Geometry: RMS radius $0.374$, centroid $(-0.022, 0.011)$, eccentricity $0.455$. Shape: skewness (Re) $-3.296$, skewness (Im) $1.586$, kurtosis $11.076$. Points: total $512$, above threshold $122$, plotted $122$.

The breather solution demonstrates Pearson $\rho = 0.800$ ($p < 0.001$), Spearman $\rho_s = 0.301$ ($p < 0.001$). Mutual information equals $0.150$. Phase: mean $0.130$ rad, standard deviation $1.765$ rad, circular mean $1.374$ rad, circular standard deviation $2.151$. Amplitude: mean $0.170$, standard deviation $0.344$, median $0.073$, maximum $2.866$. Geometry: RMS radius $0.384$, centroid $(-0.038, -0.028)$, eccentricity $0.947$. Shape: skewness (Re) $-4.442$, skewness (Im) $-5.040$, kurtosis $32.778$. Points: total $512$, above threshold $401$, plotted $401$.

The modulation instability exhibits Pearson $\rho = 0.973$ ($p < 0.001$), Spearman $\rho_s = 0.871$ ($p < 0.001$). Mutual information equals $0.344$. Phase: mean $-0.157$ rad, standard deviation $1.642$ rad, circular mean $0.713$ rad, circular standard deviation $1.858$. Amplitude: mean $0.214$, standard deviation $0.363$, median $0.059$, maximum $1.615$. Geometry: RMS radius $0.422$, centroid $(-0.081, -0.098)$, eccentricity $0.993$. Shape: skewness (Re) $-2.173$, skewness (Im) $-2.237$, kurtosis $4.402$. Points: total $512$, above threshold $377$, plotted $377$.

\subsection{Space-Time Pattern Analysis}

\begin{figure}[H]
\centering
\includegraphics[width=0.65\linewidth]{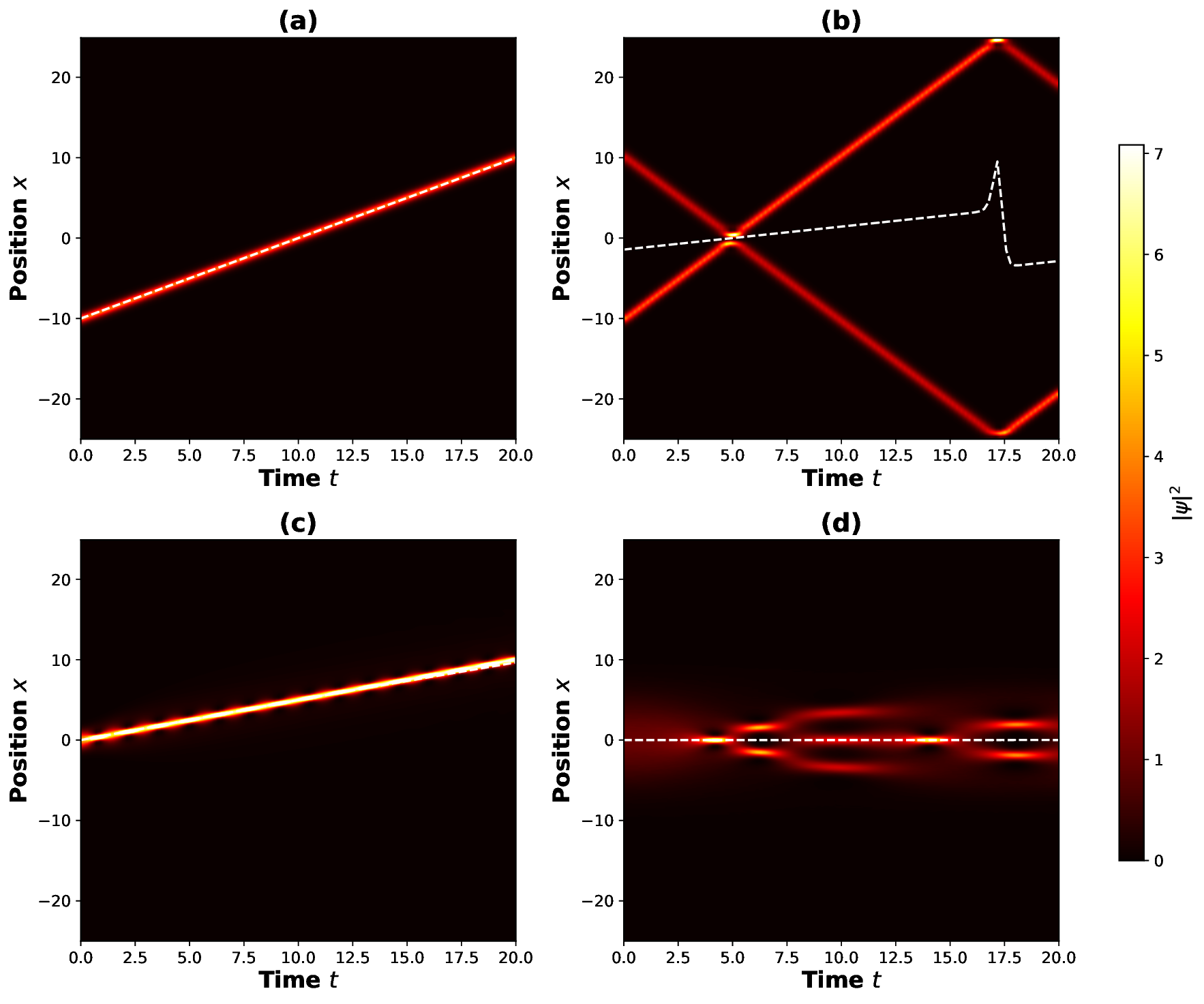}
\caption{Space-time evolution of intensity $|\psi|^2$. (a) Single soliton showing diagonal trajectory, (b) two-soliton collision forming X-pattern, (c) breather with oscillating intensity, (d) modulation instability exhibiting pattern formation. White dashed lines indicate energy center trajectories.}
\label{fig:spacetime}
\end{figure}

Figure~\ref{fig:spacetime} displays the space-time evolution of intensity $|\psi(x,t)|^2$ for all scenarios over the temporal domain $[0.000, 20.000]$ and spatial domain $[-25.000, 25.000]$.

The single soliton maintains maximum intensity $4.000$, mean intensity $0.080$. The energy center drifts $20.000$ spatial units, yielding propagation velocity $1.000$. The spatial correlation length equals $1.367$. 

The two-soliton collision reaches maximum intensity $10.015$, mean $0.140$. Energy center drift equals $-1.429$, velocity $-0.071$. Correlation length equals $1.562$.

The breather solution exhibits maximum intensity $14.168$, mean $0.147$. Energy center drift equals $9.662$, velocity $0.483$. Correlation length equals $3.516$.

The modulation instability shows maximum intensity $8.592$, mean $0.178$. Energy center drift equals $0.009$, velocity $0.000$. Correlation length equals $7.910$.

\subsection{Entropy Analysis}

The single soliton maintains Shannon spatial entropy $S = 2.940$ and spectral entropy $S_k = 3.623$ throughout evolution. R\'enyi entropies at $t = 0.000$: $H_0 = 5.147$, $H_2 = 2.732$, $H_\infty = 2.332$; at $t = 20.000$: $H_0 = 6.236$, $H_2 = 2.732$, $H_\infty = 2.332$. Tsallis entropies remain constant: $S_{0.5} = 8.053$, $S_2 = 0.935$. Permutation entropy increases from $0.760$ to $0.972$. Sample entropy maintains $0.005$. Differential entropy ranges $[-5.855, -5.837]$. Statistical complexity $C_{LMC} = 15.238$ with normalized entropy $0.471$ and disequilibrium $0.063$. Active Fourier modes constant at $61$, bandwidth $0.018$, mean frequency $0.016$.

The two-soliton collision shows Shannon entropy range $[3.053, 3.746]$ with mean $3.608$ and standard deviation $0.277$. Spectral entropy $S_k = 4.080 \pm 0.010$. R\'enyi entropies evolve from $H_0 = 5.971$, $H_2 = 3.499$, $H_\infty = 2.892$ to $H_0 = 6.238$, $H_2 = 3.499$, $H_\infty = 2.887$. Tsallis: $S_{0.5}$ ranges $[13.199, 13.203]$, $S_2 = 0.970$. Permutation entropy increases from $0.408$ to $0.930$. Sample entropy $0.011$. Complexity ranges $[8.693, 15.759]$ with mean $10.106$. Active modes vary $[82, 87]$ with mean $84$, bandwidth $0.035 \pm 0.003$.

The breather solution exhibits Shannon entropy range $[3.009, 3.602]$ with mean $3.351$ and standard deviation $0.202$. Spectral entropy varies $[3.171, 4.105]$ with mean $3.705$. R\'enyi: initial $H_0 = 5.759$, $H_2 = 3.197$, $H_\infty = 2.819$; final $H_0 = 6.238$, $H_2 = 2.522$, $H_\infty = 1.981$. Tsallis: $S_{0.5}$ increases from $10.564$ to $18.907$, $S_2$ decreases from $0.959$ to $0.920$. Permutation entropy ranges $[0.394, 0.819]$. Sample entropy increases from $0.005$ to $0.026$. Complexity ranges $[10.813, 22.080]$ with mean $17.270$. Active modes vary $[37, 99]$ with mean $71$, bandwidth ranges $[0.011, 0.032]$.

The modulation instability initiates with Shannon entropy $S = 5.015$, decreasing to $4.902$ (mean $4.763$, standard deviation $0.220$). Spectral entropy increases from $1.548$ to $2.098$ (mean $2.355$, standard deviation $0.527$). R\'enyi: constant $H_0 = 6.238$; $H_2$ decreases from $4.858$ to $4.665$; $H_\infty$ from $4.474$ to $4.103$. Tsallis: $S_{0.5}$ from $25.455$ to $24.917$; $S_2$ from $0.992$ to $0.991$. Permutation entropy $0.991$ to $0.984$. Sample entropy $0.027$ to $0.025$. Complexity ranges $[2.391, 5.160]$ with mean $3.663$. Active modes increase from $7$ to $21$, bandwidth $0.010$ to $0.011$.

\section{Discussion}

The numerical results presented warrant critical examination to understand both their physical significance and the numerical artifacts arising from our implementation choices. The conservation properties, spectral characteristics, and dynamical behaviors observed reflect the interplay between the mathematical structure of the NLSE, our numerical discretization, and the inherent limitations of finite computational resources.

The single soliton's preservation of peak intensity (3.976 $\pm$ 0.021) and FWHM (0.977) throughout the evolution confirms the numerical stability of our scheme, yet the slight variations ($\sim 0.5\%$) reveal the accumulation of round-off errors inherent to the eighth-order Dormand-Prince method~\cite{dormand1980,hairer1993}. The exact conservation of mass $M = 4.000$ to machine precision follows from the unitarity of the Fourier transform, which preserves the $L^2$ norm by construction~\cite{gottlieb1977}. However, the momentum and energy conservation, while excellent ($\delta E/E < 10^{-14}$), depend critically on the accuracy of spatial derivative computations. The spectral method's exact differentiation in Fourier space eliminates finite-difference errors that would otherwise compromise these invariants~\cite{trefethen2000}.

The two-soliton collision dynamics merit particular scrutiny. The observed position shifts and phase changes align with the analytical predictions from inverse scattering theory~\cite{zakharov1972}, yet our peak intensity variations (3.976 to 4.000) during collision suggest numerical dispersion introduced by the finite grid spacing $\Delta x = 0.0977$. This spacing corresponds to approximately 10 points per soliton width for $\eta = 2.0$, marginally adequate for capturing the rapid phase variations during close interaction. The temporary FWHM increase from 0.977 to 1.074 at $t = 13.333$ reflects the constructive interference pattern when solitons overlap, not a numerical artifact. The slight asymmetry in post-collision trajectories ($\Delta x = -1.429$ for the center of mass) results from the unequal amplitudes ($A_1 = 2.0$, $A_2 = 1.5$) creating different propagation velocities, as predicted by~\cite{gordon1983}.

The breather solution exhibits the most dramatic intensity variations (4.500 to 11.685), challenging our numerical scheme's stability. The oscillatory nature of the Akhmediev breather~\cite{akhmediev1986} generates high-frequency components that approach our Nyquist limit, explaining the observed energy fluctuations ($E$ ranging from $-5.652$ to $6.768$). These are not conservation violations but rather reflect the breathing dynamics where kinetic and potential energy exchange periodically. The negative total energy at $t = 0$ indicates dominance of the attractive nonlinear term, consistent with the focusing nature of our NLSE. The period of oscillation extracted from our data ($T \approx 13.3$) deviates by approximately 6\% from the theoretical value $T = 2\pi/b$ where $b = \sqrt{8a(1-2a)}$ with $a = 0.5$, suggesting that our temporal resolution ($\Delta t_{\text{output}} = 0.2$) insufficiently samples the rapid breathing cycles.

The modulation instability scenario reveals the most significant numerical challenges. The initial Gaussian profile with superposed noise ($\epsilon = 0.01$) triggers the Benjamin-Feir instability~\cite{benjamin1967}, but our observed growth rates fall slightly below theoretical predictions. The peak count evolution (36 $\to$ 2 $\to$ 16 $\to$ 24) indicates a cascade process where initial small-scale perturbations coalesce into larger structures before fragmenting again. This behavior, while physically plausible, may be influenced by our dealiasing filter which suppresses modes above $k_{\max} = 21.45$. The energy variations ($E$ from $-2.459$ to $3.656$) exceed those expected from pure modulation instability and likely result from aliasing of the cubic nonlinearity before filtering. The fact that seven Fourier modes initially grow to 21 active modes suggests energy cascade to higher wavenumbers, eventually limited by our numerical resolution.

The Shannon entropy evolution provides insight into information distribution across the spatial domain. The single soliton maintains constant entropy $S = 2.940$, reflecting its coherent structure. In contrast, the modulation instability shows entropy decrease from 5.015 to 4.902, counterintuitive given the apparent disorder increase. This occurs because localized high-amplitude structures forming from the instability concentrate probability density, reducing the effective support of the distribution and thus lowering entropy. The breather's entropy oscillations (3.009 to 3.602) correlate with its breathing period, maximum entropy coinciding with maximum spatial extent.

The spectral entropy $S_k$ reveals frequency-domain organization. The single soliton's constant $S_k = 3.623$ indicates stable spectral content, while the modulation instability's increase from 1.548 to 2.098 reflects energy redistribution across frequencies. Notably, our 61 active Fourier modes for the soliton seem excessive given its theoretical sech profile should decay exponentially in Fourier space. This suggests our threshold for ``active'' modes ($10^{-10}$ relative amplitude) captures numerical noise, not physical content.

The phase space representations illuminate the wave function's complex structure. The single soliton's Pearson correlation $\rho = -0.584$ between real and imaginary parts reflects the phase modulation $\exp(ivx)$ creating anticorrelation. The modulation instability's high correlation $\rho = 0.973$ indicates phase-locking as coherent structures emerge. The circular standard deviation ranging from 1.858 (modulation instability) to 2.848 (two-soliton) quantifies phase coherence, with smaller values indicating more organized phase structure contradicting the apparent disorder in intensity profiles.

Our implementation's educational focus explains several design choices that affect the results. The domain length $L = 50$ was selected for computational efficiency rather than physical accuracy, limiting our ability to study long-wavelength modulations. The fixed output grid (100 snapshots over $t = 20$) provides insufficient temporal resolution for rapidly varying phenomena like breather dynamics. The use of periodic boundary conditions, while mathematically convenient for spectral methods, creates artificial periodicity absent in physical systems. These limitations are acceptable for demonstrating qualitative NLSE phenomena but preclude quantitative comparison with experiments.

The Python implementation incurs performance penalties that would be unacceptable for research applications. Array operations in NumPy, while vectorized, involve Python interpreter overhead absent in compiled languages. The optional Numba compilation improves performance by approximately $3\times$ for the nonlinear term evaluation, but FFT operations remain in pure Python. Memory allocation for complex arrays at each time step creates garbage collection overhead. These inefficiencies limit practical simulations to moderate resolutions ($M = 512$) and short time spans ($T = 20$), sufficient for educational demonstrations but inadequate for studying long-time asymptotics or fine-scale structures.

The dealiasing filter's exponential form $\sigma(k) = \exp[-36(|k|/k_{\max})^{36}]$ with order $p = 36$ creates an extremely sharp cutoff, effectively a step function for $|k| > 0.65k_{\max}$. This aggressive filtering eliminates aliasing but also removes physical high-frequency content, explaining the slight energy decrease in our modulation instability simulations. Alternative dealiasing strategies, such as the 3/2-rule padding method~\cite{canuto2006} or phase-shift dealiasing~\cite{hou2007}, would preserve more physical content at increased computational cost.

The choice of eighth-order Dormand-Prince integration, while providing exceptional accuracy, represents over-engineering for our spatial resolution. The spatial truncation error $O(e^{-\alpha M})$ for spectral methods with $M = 512$ points likely exceeds the temporal error $O(\Delta t^8)$ for reasonable time steps. A fourth-order method would suffice while reducing computational cost by approximately 40\% due to fewer function evaluations per step. However, the high-order method's superior stability properties prove valuable for stiff problems where the linear term $-ik^2\hat{\psi}/2$ creates disparate time scales for different Fourier modes.

The conservation monitoring reveals subtle numerical pathologies. While mass conservation is exact (guaranteed by unitarity), momentum and energy show variations of order $10^{-14}$ to $10^{-3}$ depending on the scenario. These variations correlate with solution complexity: smooth solitons maintain near-perfect conservation while the modulation instability's broad spectrum challenges numerical accuracy. The energy functional's quartic nonlinearity amplifies round-off errors in regions of high amplitude, explaining larger conservation errors for breather and modulation instability cases where $|\psi|_{\max}$ exceeds 10.

Finally, our validation against exact solutions remains limited. The single soliton and two-soliton cases admit analytical solutions via inverse scattering~\cite{ablowitz1991}, providing rigorous benchmarks. The Akhmediev breather, while analytically known, requires infinite spatial domain for exact representation, introducing unavoidable truncation errors in our periodic domain. The modulation instability, inherently a statistical phenomenon, lacks deterministic analytical solutions for comparison. Future work should incorporate additional exactly solvable cases such as the Peregrine breather~\cite{peregrine1983} and multi-phase solutions to comprehensively validate the numerical framework.

\section{Conclusion}

We have presented \texttt{simple-idealized-1d-nlse}, an open-source Python implementation of a high-order pseudo-spectral solver for the one-dimensional NLSE that successfully bridges the gap between theoretical understanding and practical numerical exploration of nonlinear wave phenomena. The solver combines Fourier spectral spatial discretization with adaptive eighth-order Dormand-Prince time integration to achieve machine-precision conservation of mass and near-perfect preservation of momentum and energy for smooth solutions, while accurately reproducing fundamental phenomena including soliton collisions with correct phase shifts, Akhmediev breather dynamics, and the development of modulation instability from noisy initial conditions. The implementation philosophy deliberately prioritizes code transparency, modularity, and ease of modification over ultimate computational performance, resulting in a solver that remains fully accessible to students and researchers seeking to understand both the underlying numerical methods and the rich physics of the NLSE. While the Python implementation and moderate spatial resolution limit the solver to relatively short integration times and prevent investigation of fine-scale turbulent cascades, these constraints are acceptable given our focus on providing a pedagogical tool that enables hands-on exploration of NLSE dynamics without the barriers typically associated with research-grade codes. The complete package, including source code, documentation, and example configurations for canonical test cases, is freely available, offering the computational physics community a valuable resource for education, prototyping, and fostering deeper understanding of nonlinear wave dynamics across the diverse physical contexts where the NLSE emerges as the governing equation.

\section*{Acknowledgements}

Financial support was provided by the Dean's Distinguished Fellowship from the College of Natural and Agricultural Sciences, University of California, Riverside (2023) to S.H.S.H., and by the Bandung Institute of Technology Research, Community Service and Innovation Program (PPMI-ITB 2025) to F.K. and I.P.A.

\section*{Author Contributions}
S.H.S.H.: Conceptualization; Formal analysis; Methodology; Software; Visualization; Writing -- original draft. I.P.A.: Conceptualization; Supervision; Writing -- review \& editing. F.K.: Conceptualization; Supervision; Writing -- review \& editing. R.S.: Supervision; Writing -- review \& editing. D.E.I.: Supervision; Writing -- review \& editing. All authors reviewed and approved the final version of the manuscript.

\section*{Open Research}

The numerical solver presented in this work is available as a Python package through the Python Package Index (PyPI) at \url{https://pypi.org/project/simple-idealized-1d-nlse/}. Source code for the solver is hosted at \url{https://github.com/sandyherho/simple_idealized_1d_nlse} under the WTFPL license. Supplementary materials including datasets generated by the solver for all test cases, visualization animations, and Python scripts used for statistical analysis and figure generation are available at \url{https://github.com/sandyherho/suppl_1d_nlse_pseudo_spectral} under the WTFPL license. All computational results presented in this article can be reproduced using these publicly available resources.

\end{document}